# Gain As Configurable Disorder: Adaptive Pumping for Control of Multimode Fiber Amplifiers and Lasers

**Tom Sperber[1,2], Vincent Billault[2], Bernard Dussardier[4], Sylvain Gigan[2], Patrick Sebbah[1,3]**

*Correspondence: Tom Sperber\* sperbertom@outlook.com ; vincent.billault@thalesgroup.com ;*

*bernard.dussardier@inphyni.cnrs.fr ; sylvain.gigan@lkb.ens.fr ; patrick.sebbah@biu.ac.il*

[1] *ESPCI ParisTech, PSL Research University, CNRS, Institut Langevin, 1 rue Jussieu, F-75005, Paris, France*

[2] *Laboratoire Kastler Brossel, ENS-PSL Research University, CNRS, UPMC-Sorbonne universités, Collège de France ; 24 rue Lhomond, F-75005 Paris, France*

[3] *Department of Physics, Jack and Pearl Resnick Institute for Advanced Technology, Bar-Ilan University, Ramat-Gan, 5290002 Israel*

[4] *Université Nice-Sophia Antipolis, CNRS, Laboratoire de Physique de la Matière Condensée, UMR7336, Parc Valrose, 06108 Nice, France*

**Abstract**

Media featuring both optical gain and disorder, such as random lasers, represent formidable challenges as subjects of research due to the high complexity of the light propagation within them; however, dramatic advances in this nascent field have been furnished by the paradigm of applying wavefront shaping techniques to the beam pumping the system. We present here a theoretical and experimental study employing this approach in a gain medium where the disorder arises not from random scattering but rather from the multimodality of a waveguide: an amplifying multimode fiber. The shaping of the core-guided pump prior to its injection affects the complex, speckle-like patterns of excitation within the fiber volume. Thus we are offered the intriguing prospect of manipulating the spatially heterogeneous gain in a system where the disorder, albeit highly complex, may be fully understood in terms of the discrete eigenmodes of a well-known waveguide. We study our medium in two different configurations: as an optical amplifier, and as a fiber laser. In the first configuration we show that dependence upon the pump configuration, of the amplifier's transmission function, surprisingly survives several physical mechanisms which at first sight would appear to severely limit it. This insight is then carried on to the lasing cavity configuration, where we find striking parallels between the system's behavior and that of random lasers; more importantly, the pump shaping shows a strong ability to control the complex emission, in particular - to selectively favor individual spectral lines and stabilize single-mode operation.

**Introduction**

Spatial modulation of the wavefront of a coherent beam prior to its coupling into scattering or complex media enables control of the light's propagation within them; multimode fibers (MMF) are a notable example of such media that has drawn intense interest in recent years. Most of the research applying wavefront shaping (WFS) techniques to MMF has dealt with passive fibers, i.e. fibers with no gain; there the transmission is linear and lends itself very well to study within the framework of the transmission matrix (TM), as defined in [1]. Using the TM approach and WFS schemes based on spatial light modulators (SLM), precise tailoring of the complex intensity patterns emerging at the passive MMF output has been demonstrated [2-6]. In parallel, WFS methods have also enabled exciting advances in quite a different domain related to the study of waves in complex media: that of random lasers. In these open, highly-nonlinear systems, shaping of the pump's spatial profile has shown a surprising ability to control the laser emission. Several types of systems were studied: An optofluidic random laser with a one-dimensional WFS scheme [7-8], a 2D Quantum Cascade Laser [9-10], and a semiconductor microdisk [11-13] (in the latter example, disorder arises from surface roughness rather than random scatterers). Using the fundamental paradigm of the RL research - that of harnessing the spatial multimodality of the **pump** lightwave as available degrees of freedom for control of a system with gain - we set out to explore an amplifying MMF with a spatially configurable pump, both in an amplifier configuration and in lasing cavity configuration.

Several applicative domains motivate an examination of potential control schemes of multimode fiber amplifiers (MMFA) and multimode fiber lasers. In the context of long-haul optical telecommunication links, where amplifiers are crucial in-line repeaters, the imminent exhaustion of the information capacity of singlemode links drives the development of Spatial Domain Multiplexing (SDM) schemes [14-15]. Within this field, several works exploring spatial modulation schemes for the pump have been recently carried out, with the interest focused on the requirement of "gain equalization", i.e. the compensation of the high-order modes' tendency to suffer greater losses in the long-haul link, as well as lower gain values within the amplifier [16]. Successful equalization by manipulation of the pump's modal content using phase plates [17-18] or an SLM [19], has been demonstrated. Since the number of guided signal modes relevant for the framework of SDM applications is typically 4-6 modes, the work was limited to this FMF (few mode fiber) realm. Similarly, only the absolute amplification values of the different modes were considered, i.e. not the full information relevant to observation of the amplifier output in the spatial domain. In another applicative field – that of high-power fiber sources and amplifiers for industrial applications such as marking, machining, and sensing – multimodality naturally enters the scene because of the need to use large core diameters fibers, in order to accommodate large intensities while avoiding non-linear effects [20]. The resulting increase in the number of guided modes is largely undesired, and various design approaches and measures are taken to suppress the high-order modes so as to maintain good beam quality [21-24]. In this context, spatial modulation approaches for control have been explored in recent works; notable examples are [25-28], where the spatial control of the output beam is achieved by shaping the signal beam prior to injection into the amplifier; to the best of our knowledge, shaping of a coherent pump in a high-power fiber amplifier has not yet been explored. Lastly, apart from the above-mentioned applicative motivations, the system under study is interesting from a more fundamental research perspective: the shaping has a unique quality of modifying not the light that is propagating through a given disorder, but rather the disorder itself - because the multimodal pump excites the gain medium in speckle-like patterns which act as complex and heterogeneous gain upon the emitted light. Conversely to the RL case, the process takes place in a waveguide with a discrete set of known propagation modes, thereby offering an attractive opportunity to study modifiable disorder in a 3D volume with a certain level of theoretical modeling being feasible.

As the outset of this work, we study the medium in the amplifier configuration, specifically in the non-depleted pump regime, where it may be considered linear with respect to the amplified signal's field; we propose a theoretical model for the gain-dependent TM, i.e. the transmission matrix of the fiber as a function of the modal content of the injected pump beam. Starting with a description of the propagation of light in a multimode fiber in the absence of gain or absorption, we first choose a set of transverse modes $\psi_i$, $i = 1 \dots N$ which best represent the fiber's eigenmodes. This essentially means that the modes propagate with negligible coupling to each other, changing only the phase terms of their complex amplitudes. Therefore, the optical field at any point along the fiber's axis of propagation z is written as:

$$E(z) = \sum_i^N A_i * e^{-j\beta_i z} * \psi_i (r,\theta) \qquad (1)$$

Where $r,\theta$ are the transverse coordinates, $\beta_i$ is the propagation constants of eigenmode $\psi_i$, and $A_i$ is the complex amplitude with which each eigenmode was launched into the fiber at the injection point at z=0. Stated in matricial terms, eq. (1) simply means that the TM of the fiber (expressed in the modal basis) is readily given as a diagonal matrix with the phases along the diagonal determined by the propagation velocities and the fiber length. The base used in our numerical simulations was the set of Linearly-Polarized (LP) modes well-known in the textbook literature of MMFs [29-30], however our theoretical model is hereby presented without loss of generality, and any set of orthogonal modes chosen to approximate the passive waveguide's base of eigenmodes may be plugged in. As was recently demonstrated in [6], for propagation distances on the order of centimeters, the LP modes may serve with excellent precision as such an approximation, and the set of circularly polarized (CP) modes may be used for more precise modeling of the propagation over a longer fiber.

Eq. (1) provides a fully deterministic prediction of the evolution of the optical field along the fiber, once the set of values of all complex amplitudes $A_i$ – which we refer to as the *modal content,* or *configuration* – has been specified. Numerical simulation of this evolution is readily performed by modelling the fiber as a series of truncated thin 'slices', each of a length $\Delta z$ fixed to be sufficiently short so as to finely sample the dephasing length $\Delta z < \pi/(max\{\beta_i\} - min\{\beta_i\})$. Since the eigenmode set is a base for any guided lightfields, the injection amplitudes are determined as the coefficients of decomposition upon this base, of the optical field injected at the fiber facet:

$$A_i = \iint E_{incident}(z=0) \cdot \psi_i^* dr d\theta \tag{2}$$

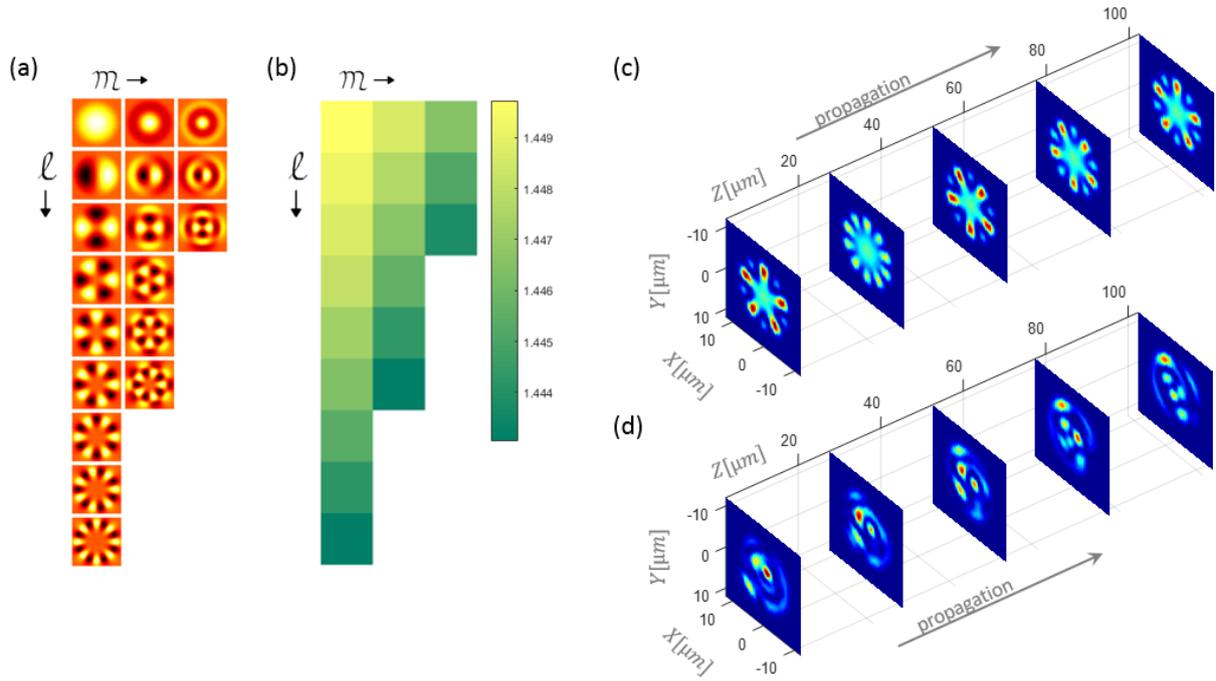

Fig. 1 – Visualization of how the modal dispersion gives rise to the seemingly disordered evolution of the intensity propagating in a MMF as complicated and changing speckle patterns. (a), (b) – the spatial field distributions and the effective refractive indices, respectively, of the 18 LP modes guided by the modeled fiber. To the right - examples of propagation, for 2 possible modal configurations, with the light's transverse intensity patterns shown in different cross-sections of the fiber. (c) - the case of only 2 modes (LP01 and LP51) excited. The propagation causes a periodic beating of these modes. (d) - same cross-sections for a case where all the modes displayed in (a) are excited.

Gain is now added to the medium in the form of a rare-earth element doping, spread throughout the fiber core with uniform doping concentration $N_d$. We follow the standard treatment in the literature, as in [31-32], where the electronic levels are modeled as a two-level gain system, with $n_1, n_2$ representing, respectively, the fractional lower and upper level populations. The 'pump' wavelength $\lambda_{pump}$ is the lightwave efficiently exciting the upper level population, and is guided by the modes $\psi_i^p$; while stimulated emission from this excitation efficiently amplifies the 'signal' wavelength $\lambda_{sig}$, with the corresponding set of modes $\psi_i^s$. Lastly, we work under the assumption of weak signal power with respect to the pump power (commonly known as non-depletion of the pump). In this regime, the rate equations are easily solved by considering only the pump intensity $I_p$ arriving at any given point:

$$n_2 = \frac{\sigma_a^{pump} \cdot I_p}{(\sigma_a^{pump} + \sigma_e^{pump}) \cdot I_p + hc/(\tau \lambda_{pump})} \tag{3}$$

Where $\sigma_a^{pump}, \sigma_e^{pump}$ are, respectively, the absorption and emission cross-sections at the pump wavelength, and $\tau$ is the lifetime of the upper level excitation.

Any optical intensity $I$ propagating through the medium is amplified/absorbed by the upper/lower levels, respectively, with efficiencies determined by the (wavelength-dependent) emission/absorption cross-sections $\sigma_e, \sigma_a$ thus:

$$\frac{dI}{dz} = (\sigma_e n_2 - \sigma_a n_1) N_d \cdot I \qquad (4)$$

Expressed in terms of field $E$ rather than intensity, the interaction with the levels may be written as a modification of the medium's refractive index, rendering it complex: $n = n_0 + n' + j \cdot n''$, where $n_0$ is the index in the absence of the gain element, $n''$ contributes the amplitude change (gain or absorption), and therefore satisfies $n'' = \frac{1}{2}(\sigma_e n_2 - \sigma_a n_1) N_d$, and $n'$ determines the associated (gain-induced) change of the wave's phase. This relation holds for both pump and signal wavelengths, if the corresponding cross-sections are plugged in. The real and the imaginary parts of the complex susceptibility are linearly related [33], and the factor relating the two parts, often referred to as the Refractive Index Change (RIC) parameter $K_{RIC} = n'/n''$, depends on the medium and its specific electronic level structure. Values may be found in the literature for specific media [34-36]. Taking into consideration eq. (3) and (4), we may express the optical field that has propagated a very short distance $\Delta z$ through the medium, as:

$$E(z + \Delta z) = E(z) \cdot e^{-j\beta_0 \Delta z} \cdot \left\{ 1 - j\beta_0 \Delta z \cdot \frac{1}{2}(\sigma_e n_2 - \sigma_a n_1) N_d \cdot (j + K_{RIC}) \right\} \qquad (5)$$

because the gain-induced changes to the refractive index are small enough to be approximated using the first-order Taylor expansion of the exponential function: $e^{x+\varepsilon} \approx e^x(1 + \varepsilon)$.

We now confront the central question of our system: how does the signal field interacting with a population excitation with some arbitrary, 'speckly' spatial distribution (because it has been generated by the multimodal pump's intensity) propagate through the fiber? It is evident from eq. (5) that once the signal field has been amplified along some given fiber slice, it is no longer necessarily a perfectly guided field; the stimulated emission contribution which it has picked up mirrors the complicated pump speckle. The approach at the heart of our model is therefore to treat the next slice as equivalent to a fiber facet upon which this field is impinging, and recouple the field perturbed by the complicated amplification back to the waveguide using the decomposition principal of eq. (2). As a result of this principle where amplification is followed by decomposition upon the mode base, the transmission matrix of each fiber slice is constructed in the following manner, element by element: the entry at some column 'c' and some row 'r', corresponding to the contribution from an incoming mode $\psi_c^s$ to an outgoing mode $\psi_r^s$, is determined as:

$$TM_{[r,c]}^{z \to z+\Delta z} = e^{-j\beta_c \Delta z} \{ \delta_{rc} + \frac{1}{2}\beta_0 \Delta z N_d (1 - jK_{RIC}) \iint [\sigma_e^{sig} n_2(r,\theta,z) - \sigma_a^{sig} n_1(r,\theta,z)] \cdot \psi_r^s(r,\theta) \cdot \psi_c^{s*}(r,\theta) \, dr d\theta \} \qquad (6)$$

where the spatial-overlap integral arises from the decomposition of the amplified field of eq. (5) upon all supported signal modes in the fiber, and the Kronecker delta replaces the unity term in the same equation so as to reflect, for the diagonal TM elements, the passive propagation of the incoming mode $\psi_c^s$.

Finally, the total gain-dependent Transmission Matrix is readily obtained by concatenating the effects of all slices, i.e. multiplying all the 'local' TMs:

$$TM_{Total}(L, Pump) = \prod_{z'=0}^{z'=L} TM^{z'} \qquad (7)$$

To summarize our theoretical model, its implementation as a numerical simulation is as follows: an input pumping configuration, i.e. some chosen modal content for the pump, is propagated from the injection point throughout the fiber slices in an iterative manner. In each slice, the level excitation caused by the arriving pump field is calculated as per eq. (3); the way this excitation acts back upon the pump is then taken into account by calculating the modified field, i.e. the one perturbed by the local absorption as per eq. (4), and lastly propagating said modified field to the next slice according to the principle of decomposition upon all supported pump modes. The resulting data, giving the upper level excitation for every point within the fiber volume, is then used to calculate the signal TM according to eq. (6) and (7).

**Results**

At the outset of the theoretical analysis, several mechanisms which inherently limit the desired sensitivity to the pump configuration become apparent. Firstly, one may observe that the pump absorption is saturable. The crucial term through which the dependence upon the pump's modal content enters the picture is the spatial heterogeneity of the gain profiles, i.e. the transverse variation of the population-inversion within the overlap integral governing eq. (6). Therefore, pump powers strong enough to saturate this term and render it homogenous, will contribute to the overall gain- but not to the gain's dependence on the pump shaping. In other words, the 'control' effect we are seeking will appear mostly where highly-contrasted gain profiles emerge, i.e. in that part of the amplifier where the pump power is not exceedingly high (yet, still high enough to induce appreciable gain). This is illustrated in fig. 2, showing typical cases of the three regimes: saturation with low dependence upon modal content, emergence of spatial non-uniformity where the desired 'control' may take effect, and decay of the excitation with the absorption of practically all pump power.

An interesting quality of this limitation is its determination by the choice of the gain element, and more specifically - by the ratio between the emission cross-section for the signal and that of absorption for the pump: $\sigma_e^{sig}/\sigma_a^{pump}$. The qualitative reasoning behind this term is, that the degree of control could be maximized by extending the length of amplifier where non-saturated gain profiles appear - that is, by increasing the typical length of pump absorption $1/(N_d \cdot \sigma_a^{pump})$ and thus 'slowing down' the transition between saturation and complete absorption of the pump power; while at the same time requiring strong amplification of the signal over this limited length. A non-trivial point should be noted about the inability to overcome this limitation by changing the doping concentration: Although a lower concentration would indeed lengthen the segment of heterogeneous gain, it would equally reduce the amplification per unit length. This is demonstrated in the bottom panel of fig. 2, where the differential gains per unit length are plotted. The differential gain is defined as the variation in the signal transmission brought about by modification of the pump modal content. Comparison between the evolutions for the two concentrations demonstrates that change in $N_d$ would only serve to scale the amplifier behavior along the z axis, leaving the total degree of dependence upon our shaping ratio to be fully dictated by the cross-sections characterizing the gain element.

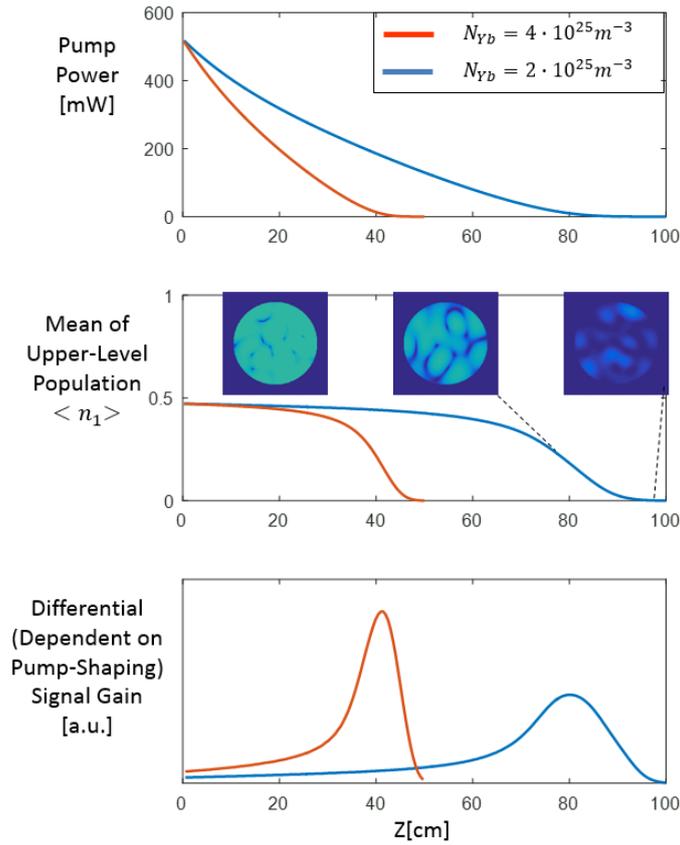

Fig. 2 – Evolution of the pump power, and the resulting medium excitation along the length of a MMFA, numerically obtained by averaging over an ensemble of 500 randomly-drawn pump configurations. Results are shown for two different doping concentration values in order to demonstrate how altering it cannot overcome the limitation upon the total accumulated 'control', and amounts only to a rescaling of the evolution in the z coordinate. Upper panel: pump power; Middle panel: upper level's excitation mean over the fiber cross-section. The insets demonstrate typical population profiles; Lower panel: the 'differential gain', as defined above.

Additional limitations stem from the fact that even in that section of the amplifier where gain profiles retain high contrast, they exhibit a complex evolution along its length. Due to the progression of the pump intensity as a series of different speckles, as predicted by eq. (1) and visually demonstrated in fig. 1, the total gain accumulated by the signal is a compound result of many realizations of disorder. Our shaping of the pump may not control this evolution, which arises naturally from the modal dispersion relations; it may only control, through the choice of the modal content, the overall statistics of the gain profiles. In short, the waveguide geometry gives rise to an effect of self-averaging of the spatial-overlap effects we wish to harness for the purpose of manipulating the signal's transmission. Lastly in the discussion of mechanisms which appear to limit the desired sensitivity, one may note that any gain profile we may choose to tailor, even at a single cross-section of the amplifier, will produce some finite gain for *all* the signal modes. This is because the interaction of the signal with the pump is an incoherent process - i.e. mediated through the scalar level excitation – and hence does not retain the pump beam's phase. In other words, no "orthogonality" exists between the degrees of freedom of our excitation (namely, the pump modes), and those of the output we wish to control (i.e. the signal modes), resulting in limited selectivity of the desired control.

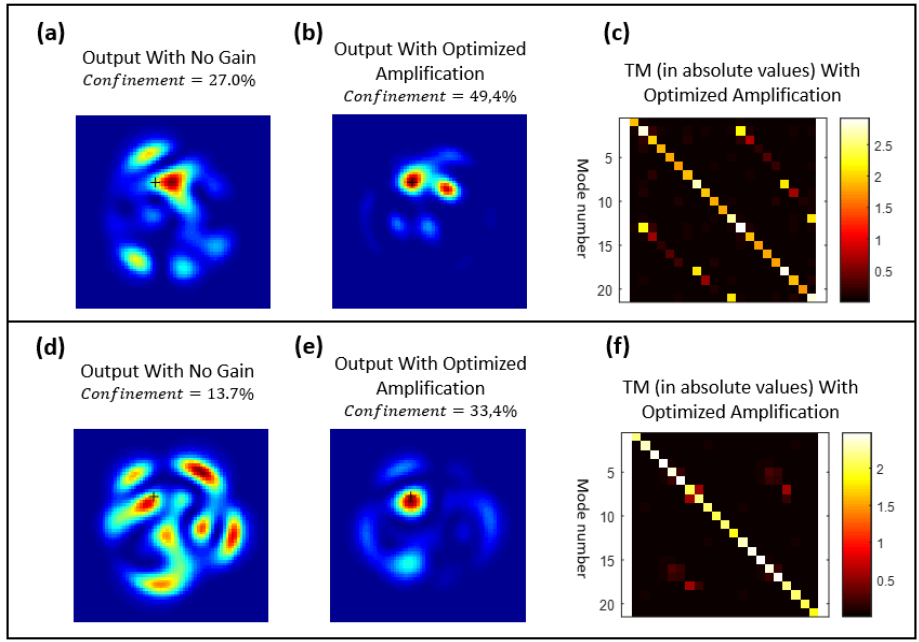

Fig. 3 – Numerical simulation of optimization the pump configuration with the goal of focusing the amplifier output intensity at the point marked by the dark + sign; results are shown for two different optimization cases, top and bottom respectively. (a) - the speckle pattern at the output in the absence of pumping; (b) - the output pattern under the pumping configuration found by the optimization function; (c) – the amplifier TM under this optimal pumping, with the colormap corresponding to the absolute values of the transmission elements. (d-f) – same as (a-c), but for the second optimization, which differs from the first only by the initial signal injected into the amplifier. The confinement metric is defined as the percentage of optical power, out of the total guided power, found in the speckle grain chosen as the focus point.

The sum total effect of these limiting mechanisms upon the desired controllability was directly evaluated by means of an optimization process of the numerically simulated MMFA. Viewed as an optimization problem, the system is conspicuously difficult to solve, containing at least several tens of degrees of freedom (twice the number of guided pump modes) which interact in a highly non-linear manner to produce the signal TM. The non-linearity ensues both from the saturation function of the medium excitation, and from the fact that the pump's intensity, rather than its field, is the excitation's driving factor; hence the gain profile of a superposition of pump modes is by no means the sum of the gains induced by the individual modes. As we may not search for an optimal configuration by a separation of degrees-of-freedom approach, the problem was addressed using a global optimization approach. A Social Spider Algorithm, which is a type of a Genetic Algorithm [37], was used to search the pumping configuration that would best produce a focus at the amplifier's output. The optimization process was repeated several times, for different modal contents of the initial signal prior to its amplification; results for two such cases are presented in fig. 3. As can be seen, the optimization yielded improvements of almost 23% and 20%, respectively for the two cases, in the metric of the confinement of the power within the speckle grain at the focus point. Evidently the control of the output pattern is not complete; however, considerable convergence towards a focus is clearly demonstrated. The interesting conclusion one may draw - contrary to what may be at first suspected based on the above discussion of inherent limitation and averaging mechanisms – is that significant ability to tailor the transmission of the MMFA survives in the system.

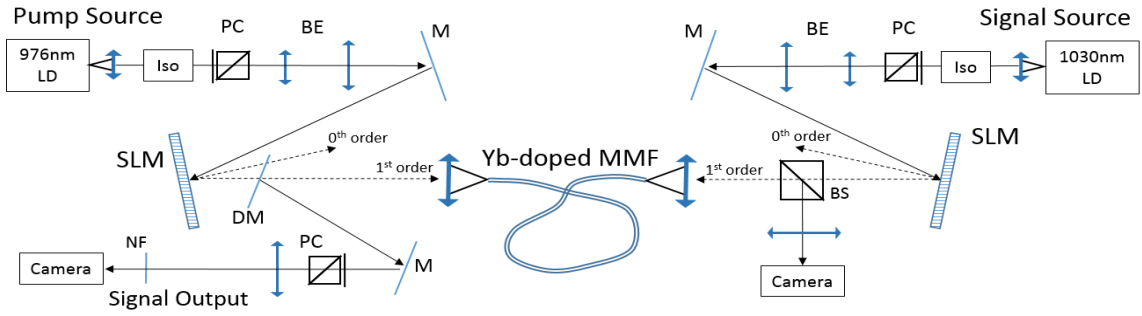

Fig. 4 – Schematic description of the amplifier experiment. Iso – Isolator, PC – Polarization Control, BE – Beam Expansion, BS – Beam Splitter, M – Mirror, DM – Dichroic Mirror, NF – Notch Filter.

   Experimental validation of our theoretical predictions is sought using the setup depicted in fig. 4, in which an Ytterbium-doped graded-index multimode fiber, guiding approximately 20 modes per polarization, serves to amplify a signal beam at 1030nm (for full details see the methods section). The experimental protocol is the search for a ***pair*** of pump configurations that show, between them, maximal difference in the transmission induced upon some fixed signal input. This search is conducted by an optimization function based on a Genetic Algorithm (GA), since – as was explained above in the section describing the numerically simulated optimization - the highly non-linear manner in which our degrees-of-control interact requires an efficient global approach. Optimizations were carried out for two types of fixed inputs: Firstly, the signal beam was randomly injected into the amplifier, producing speckle patterns at the output, and the metric maximized by the pump optimization was the *decorrelation* of the pattern. In the second type, the signal SLM was used to focus the signal through the gain, and the metric maximized by the pump optimization process was *defocusing*, quantified as the loss of confinement of the optical power within the focus spot.

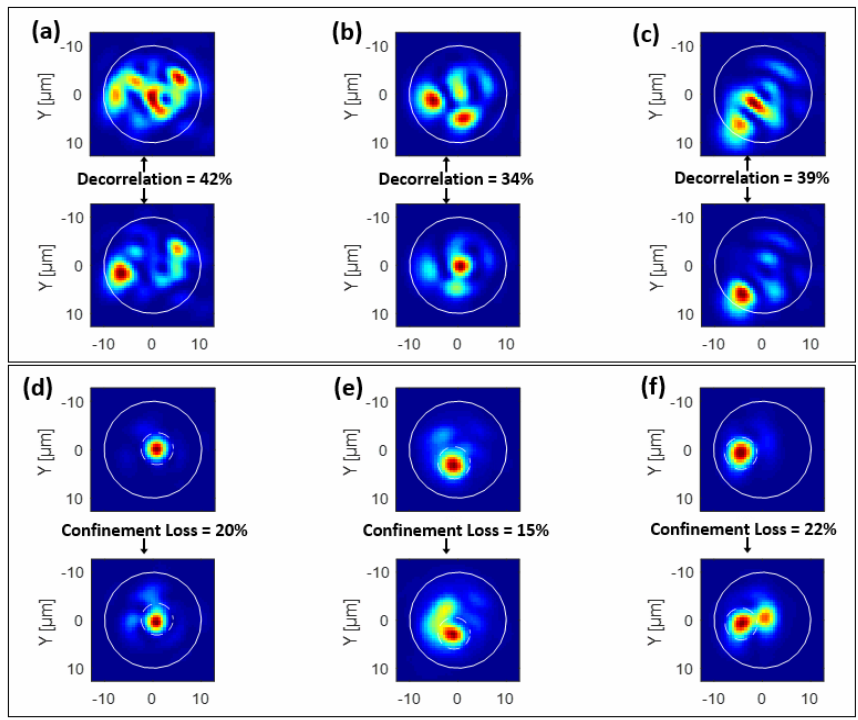

Fig. 5 – Results of the amplifier control experiment. (a) - Scatter plot of signal output decorrelation vs. the total gain, illustrating the issue of an overall baseline effect due to the presence of gain, as opposed to the variation caused purely by the choice of pump configuration. (b-d) Three instances of the experiment of signal speckle decorrelation by optimization of the pump wavefront. (e-g) Three instances of the experiment of signal defocusing by optimization of the pump wavefront; the confinement scores for the three pairs are, respectively, 68%/48%, 45%/30%, and 62%/40%. Solid white lines in the camera images delimit the fiber core; dotted white lines delimit the focus point disk over which the confinement metric is measured.

Results of optimizations where maximal decorrelation of a speckle pattern was pursued are shown in panels (b-d) of fig. 2, each panel displaying the pair of output patterns found, and the value of decorrelation – 42%, 34%, and 39%, respectively. Results of optimizations where maximal defocusing of a focal spot was pursued are similarly shown in panels (e-g) of the same figure, exhibiting losses of focus confinement of 20%, 15%, and 21.5%, respectively. The valuable outcome of the above results is the qualitative affirmation, by a relatively simple experimental setup, of the main message of the theoretical model: that non-negligible dependence of an amplifier's TM upon the pumping configuration manages, surprisingly, to survive several strong limiting and self-averaging effects. It should be readily conceded that full quantitative comparison to the numerical model's predictions remained out of scope for this implementation: For one, full measurement of the TM was not attempted, and secondly, the constraint of working at a constant gain point somewhat limited our ability to explore the entirety of the space of possible pump configurations. As an outlook for possible future work, we note that fulfillment of the first point, i.e. full knowledge of the TM, would require holographical phase measurement at the output joined with complex fitting to the real modal basis, e.g. as done in [5-6]; and that the second point may perhaps be partially addressed with more complicated WFS schemes where diffraction losses are minimized, e.g. as in [38]. The goal of our experiment, however, was not to embark on such a thorough quantitative testing of the degree to which the TM elements may be controlled, but rather to provide evidence of the interesting prediction that the overall transmission may be significantly modified by choice of the pump modal content. This highly non-trivial insight established a strong starting-point and motivation for the study of the same gain medium in a freely lasing configuration. The sensitivity of the signal's transmission function, to the pump's modal composition, suggested that an intriguing possibility to influence the dynamics of a lasing system could be offered by the same pump-shaping paradigm. As described in the following section, investigation of this prospect was tackled by a direct experimental approach.

The experimental setup for a lasing system, depicted in fig. 6, uses the same gain medium, and the same pumping scheme along with its WFS module, as in the above-described amplifier setup. The main difference is the removal of the external signal branch and the formation, around the fiber, of a cavity able to generate its own emission by oscillation. Furthermore, at the system output branch, a spectrometer is added, enabling characterization of the emission in the spectral domain, in parallel to the spatial domain imaging. The first inset in fig. 6 shows the characteristic emissions at the system output. Starting with low-power pumping, incoherent ASE appears over a broad spectrum, entirely filling the core area and the numerical aperture. Raising the pump power leads to the sudden and unmistakable appearance of oscillation, whose coherent nature is evidenced by very narrow lines in the spectrum. Typically, a few lines coexist, emerging at wavelengths with no particular 'ordered' structure, i.e. not as a periodic comb. The specific wavelengths are highly sensitive to even minor tweaking of the cavity geometry – i.e. to the placement of the fiber and the alignment of the mirrors - and seem to emerge potentially anywhere within the gain medium's broad range. In the spatial domain, the emission manifests as well-defined intensity patterns, and very interestingly exhibits increased non-uniformity and complexity when the mirrors defining the cavity are slightly tilted or defocused. The development of speckle-like patterns with increasing misalignment indicates that higher orders of the fiber's guided modes participate in the lasing – a trend directly opposed to that found in free space lasing cavities, where misalignment tends to eliminate the higher order modes.

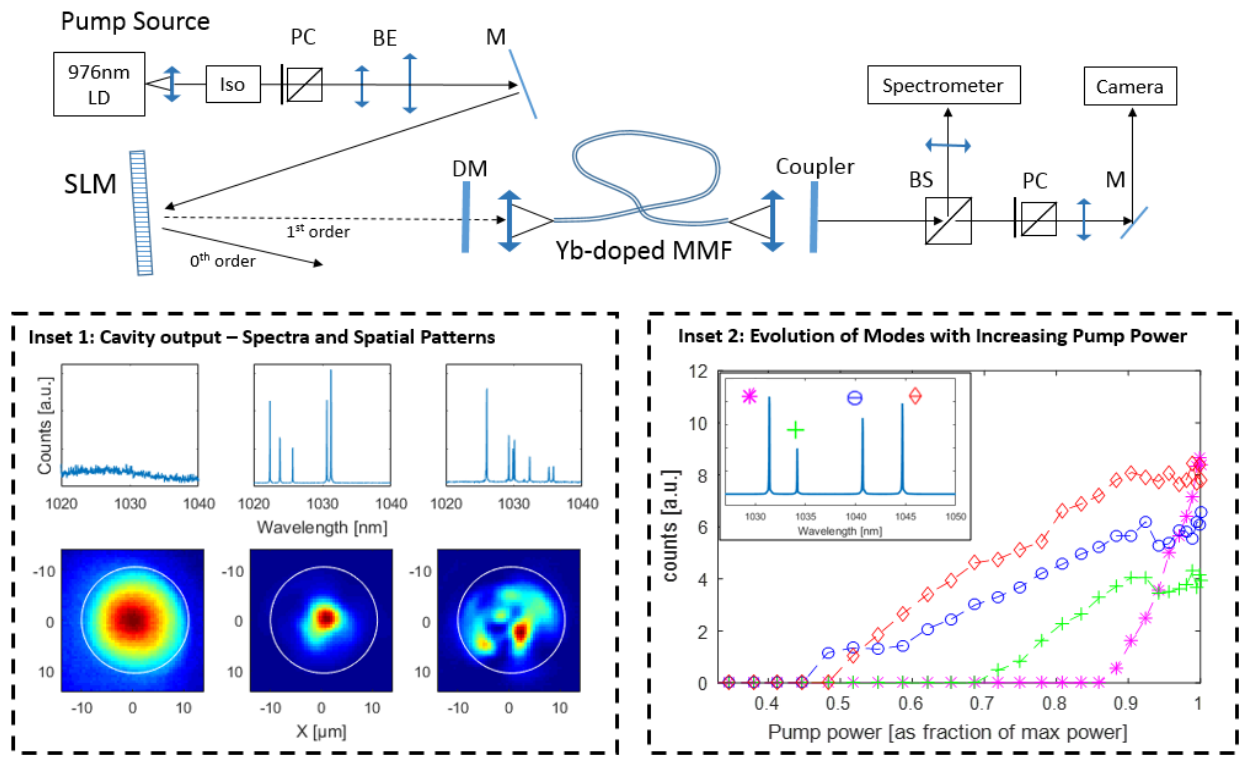

Fig. 6 – Schematic description of the Laser experimental setup, with examples of the characteristic output emission. Iso – Isolator, PC – Polarization Control, BE – Beam Expansion, BS – Beam Splitter, M – Mirror, DM – Dichroic Mirror. Inset 1 – spectra and spatial intensity patterns; left - low-power pumping (below lasing threshold); center - high-power pumping (above threshold) with the cavity's free-space optics well aligned; right - same pumping power as center, with the free-space optics deliberately misaligned. Inset 2 – An example of the modal powers as a function of the pumping power. The 4 competing modes are identified by the markers as displayed over the spectrum at the top left corner, which is the output spectrum at maximal pump power.

We may readily interpret this somewhat counter-intuitive result by recognizing the crucial *mode-mixing* role played by the coupling optics: only in an idealized system where the free space components were to perform perfect self-imaging of the fiber facet unto itself, would a LP mode travelling in the fiber also constitute an eigenmode of the entire cavity. In the real system, where the field of any outgoing fiber mode is imperfectly reproduced as an injected field, excitation of other supported modes inevitably occurs as the light is recoupled to the fiber, thereby forcing any mode of the full round trip in the cavity to be composed of some *combination* of several LP modes. Lastly, it should be noted that a measurement of the power of each spectral mode (i.e. lasing line) as a function of the input pump power reveals modal competition. An example where 4 modes compete is shown in the second inset of fig. 6: The first mode to begin oscillating (denoted by blue circles) suffers a reduction of its slope once the second mode (in red squares) starts lasing, which itself exhibits a similar reduction when the third mode (in green plus signs) appears although the first mode remains, this time, unperturbed. Lastly, when a fourth mode (in magenta asterisk signs) appears, it seems to seize the pump gain to the point of changing all previously-lasing modes to negative slopes, allowing it to eventually surpass them in power. To conclude the description of the characteristic emission, we may point to a striking resemblance of our lasing cavity's behavior to that of Random Lasers - both in the spatially complex intensity patterns and in the complicated behavior in the spectral domain – despite the lack of randomly placed scatterers within the gain medium.

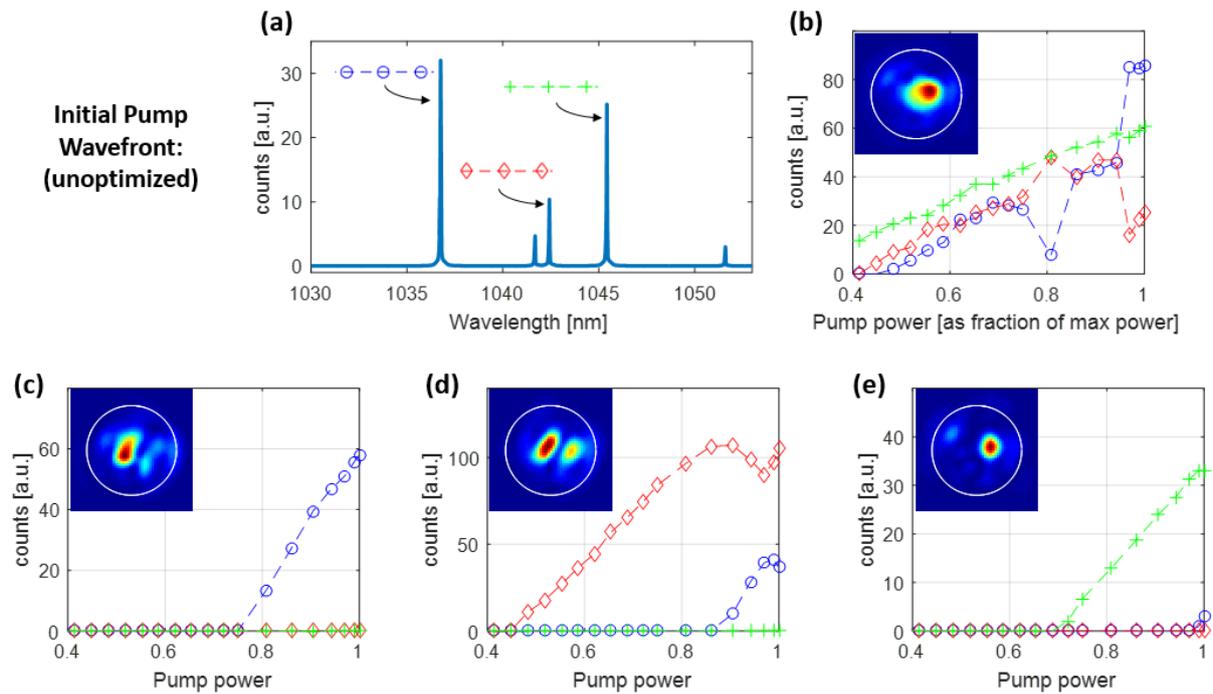

Fig. 7 - Results of the laser control experimental. The behavior of the cavity when pumped by the initial unshaped configuration is summarized in the upper half: (a) shows the spectrum measured at maximal pump power, with the color code used later to identify each of the 3 targeted modes indicated with arrows. The measured output powers for these modes vs. pump power, in the same configuration, is plotted in (b), with the inset displaying the intensity profile seen at the output facet of the fiber at maximal pump power (the white line overlay delimits the core). The results of the three optimizations directed at favoring each of the 3 modes are summarized in (c-e), with analogous color coding and inset placement as in (b).

The remarkable effect seen when spatial modulation of the pump is applied, is that for some pumping configurations, the speckle pattern at the laser output changes, accompanied by a change in the relative weights of the lasing spectral lines. This modulation effect is observed in a stable and repeatable manner – meaning the same SLM mask produces the same emission pattern - over a time-scale of tens of minutes, as long as the fiber and the cavity optics are not perturbed. In order to investigate the degree to which the modulation may control the lasing, an optimization protocol is implemented to search for pumping configurations that would maximize the relative weight, within the total power of the emission, of a chosen spectral line. The same GA function described above in the amplifier experiment is used, but this time receiving as input metric the measured spectrum, and optimizing the pump shaping without the constraint of fixed-gain configurations. Results of such optimizations, as displayed in fig. 7, strongly confirm that ability of the pump shaping to favor at will specific lasing modes. At first, the system is pumped with a default configuration - a uniform phase mask - and several co-lasing modes appear, as displayed in (a). For convenience, the three strongest ones are labeled (from left to right) by blue circles, red squares, and green plus signs - representing the modes at 1036.75nm, 1042.4nm, and 1045.4nm, respectively. The modal powers vs. pump power measurement performed in this initial configuration, shown in (b), reveals the complicated nature modal competition, similarly to the example given and discussed above (see inset 2 of fig. 6). Specifically, the 'blue' and 'red' modes seem to compete for the same gain, alternating in the roles of dominant and repressed mode, while in parallel the 'green' mode remains unaffected and increases with constant slope. Then, three separate optimizations procedures are performed, each time targeting a different mode among the aforementioned three and aiming to maximize its metric; once the GA converges to the best found SLM mask per some desired mode, measurement of the modal powers as a function of the pump power is performed under that pumping configuration. The results of these optimizations are illustrated by the graphs (c-e), which correspond, respectively, to the targeting of the 'blue', the 'red', and the 'green' mode. As can be seen for all 3 cases, a strong ability is clearly demonstrated to control the lasing behavior and selectively excite a desired mode, by the choice of the spatial modulation at the pump SLM. Furthermore, the isolation of the individual spectral modes shows, that in the spatial domain, each mode corresponds to a unique intensity pattern.

**Discussion**

We reported here on investigations into the degree of control attainable, by a scheme of spatial modulation of a coherent pump beam, of the output emission of two systems: an optical fiber amplifier, and a fiber laser, both based on the same multimode Yb-doped fiber serving as the gain medium. In the first system, a moderate ability to manipulate the intensity pattern at the output was obtained, but full control remained out of reach; this finding agreed with the predictions of our theoretical model, which clarified how several mechanisms inherent to the pump-signal interaction and to the waveguide properties impose limitations upon the desired control. Essentially, the degree of spatial overlap, summed over the entire 3D volume, between any two modal configurations of the amplified light, and between any one of these configurations and the gain distribution excited by the pump, is inevitably always quite high; hence discrimination between the two by the choice of the pump content remains difficult. By contrast, in the second system – that is, the lasing cavity - careful tailoring of the pump had a dramatic effect upon the emission, allowing robust selective control of the lasing modes strong enough to invert the ratio between their thresholds. The fact that the different lasing modes correspond to different combinations of the spatial eigenmodes of the waveguides (namely, the fiber LP modes), is evidenced by the unique intensity patterns revealed at the output when each one of the modes is isolated by the pump shaping. We may therefore see each one of the lasing modes as having a particular evolution throughout the cavity length, due to modal dispersion: of course it must constitute a standing wave in the cavity, but nonetheless one which is composed of a series of different speckle-like intensity patterns. As was the case in the amplifier system, the degree of spatial overlap between different modal configurations inevitably remains high, leading to small differences in gain; therefore, we conclude that the strong controllability found in the lasing system is a result of the presence of feedback and of the non-linear (in the sense of having a clear threshold for the onset of oscillation) dynamics, as opposed to the single-pass, linear gain of the amplifier system. Drawing also on the striking similarity in the behavior of the fiber laser to that of random lasers, we suggest that the mechanisms which explain the surprising controllability found in RL play an important role in our system as well. More specifically, in RL it has been shown that mode selection by means of pump shaping heavily relies on the nonlinear cross-coupling modal interactions, which allow small variations in the gain profile to strongly influence the modal competition and determine its outcome, even for modes which are spatially extended (weakly localized) and hence have high overlap. The study of this mechanism seems therefore to be a pertinent path forward for further work seeking to better understand the dynamics of multimode fiber-based lasing systems.

To summarize, the work hereby presented offers insights into both the possibilities and limitations of the paradigm of spatial modulation of the pump, for the control of amplifiers and lasers based on multimode fibers. The amplifier configuration was investigated both by theoretical modeling and by experimentally validating the predicted dependence upon the shaping; then, the results and insights obtained served as a useful platform for an experimental study of a lasing cavity based on the same Yb-doped fiber. For the first time, to the best our knowledge, control of the emission of a multimode fiber laser by wavefront shaping the pump is demonstrated. In parallel to the applicative interest in such systems, the study suggests an attractive avenue for future research of random lasing phenomena. In particular, behavior as complex as that found in 'classical' RL arises within a simple waveguide geometry solely from the multimodality of the guidance, and robust mode selectivity is readily achievable using the pump-shaping approach.

**Material and Methods**

The amplifier is pumped by a coherent beam at 976nm, injected in the counter-propagating direction after passing its own WFS module. The signal is supplied by a single frequency laser diode emitting at 1030nm (DFB-1030-PM-50, manufactured by Innolume), and the pump is supplied by a diode delivering 500mW at 976nm (Single Frequency fiber-coupled BF package, manufactured by IPS). In each WFS module, the beam is collimated and expanded so as to uniformly illuminate the surface of a Pluto 'Holoeye' phase-only SLM, with a pixel pitch of 8μm. The chosen modulation scheme is the 'off-axis' approach, meaning the only the light reflected at the $1^{st}$ diffraction order of a grating displayed on the slightly misaligned SLM is coupled into the fiber, as in for example [2-3]. Although this choice leads to considerable loss of optical power, it holds the important advantage of avoiding having the non-modulated $0^{th}$ order reflection as a fixed and non-controllable component in the coupled lightwave. The spatially modulated pump and signal lightwaves are focused into the amplifier, each on its respective side, by aspheric coupling lenses. Specifically in the pump branch, the beam passes through a dichroic mirror (DM) with a cut-off wavelength of 1 μm, whose purpose is to serve as a reflector for the longer wavelength output, i.e. the amplified signal, thus enabling its separation from the pump beam. This output is then sent through an additional spectral notch filter, in order to suppress most of the amplifier's incoherent spontaneous emission, and imaged upon a CMOS camera. The fiber itself has a core diameter of 18μm and a numerical aperture of 0.21; based on the measured index profile and standard MMF theory [39], the number of guided modes at wavelength ~1μm is calculated to be approximately 20 modes per polarization. The length of fiber used is 3m, significantly longer than what is needed for full absorption of the available pump power; the surplus length of un-pumped fiber is needed in our setup to add loss at the signal wavelength and thus suppress parasitic lasing, only because our fiber facets were prepared with a simple perpendicular cleave; In general this obvious penalty upon the total gain may be readily avoided by angle-cleaving. Regarding the pair-searching optimization process, we note that the goal is to study the dependence of the system's transmission upon, and only upon, the modal composition of the pump. In order to avoid the pitfall of confounding this effect with different undesired ones, the optimization process is constrained to search for pumping configurations around a constant-gain working point. This challenge arises because our spatial modulation scheme does not, in general, couple a constant pump power to the fiber. The fact that a change in the gain's total magnitude will modify the signal transmission, regardless of the spatial distribution of the pump, may be readily seen just by increasing the pump power for a given shaping; decorrelation of the output then occurs due to the increase in the gain differences between the signal modes, as well as from thermal effects..

In the lasing experiment, the cavity is formed with the help of the same DM mentioned above, but, instead of separating the signal from the pump, it is used to recouple emission back into the gain fiber. Specifically, the DM is placed at the front focal plane of the aspheric coupling lens, thus forming a 4f system folded upon itself that reproduces the outgoing light field from the fiber as a re-entering one. On the opposite side of the fiber, an identical self-imaging system is formed with a coupler mirror, through which the lasing output is collected. The coupler mirror exhibits 96% reflectance at the emission wavelength range.

**Acknowledgements**

This research was supported by the Agence Nationale de la Recherche N°ANR: 12-BS09-003-015. P. S. is thankful to the CNRS support under grant PICS-ALAMO. This research was also supported in part by The Israel Science Foundation (Grants No. 1871/15 and 2074/15) and the United States-Israel Binational Science Foundation NSF/BSF (Grant No. 2015694). The authors wish to thank Raphael Florentin, Vincent Kermene, Alain Barthélémy, and Agnes Desfarges-Berthelemot, of the XLIM research institute, and Jean-Pierre Huignard of the Langevin Institue, for their support and assistance.

## Conflicts Of Interest

The authors declare no conflicts of interest.